\documentclass[
    ,final            % use final for the camera ready runs
  ]
  {aipproc}

\layoutstyle{6x9}

\begin{document}

\title{The Synergy between Deep X-ray and Infrared Surveys: AGN and
  Star Formation Activity}


\author{J. R. Mullaney}{
  address={University of Durham, U. K.}
}

\author{D. M. Alexander}{
  address={University of Durham, U. K.}
}

% \author{<author3>}{
%   address={<common address for author2 and author3>}
%   ,altaddress={<author1 address>} % additional visiting address
% }

\begin{abstract}
  We explore the connections between the infrared and X-ray properties
  of AGNs.  Using the well constrained infrared SEDs of a sample of
  local (i.e., $z < 0.1$) sample of X-ray AGNs, we develop new
  diagnostics that exploit 24~${\rm \mu m}$ and 70~${\rm \mu m}$ flux
  densities to identify AGN-dominated systems at $z<1.5$ and measure
  their total infrared luminosities.  We apply these diagnostics to the
  X-ray detected AGNs in the Chandra Deep Field South and, in doing
  so, reveal that the infrared to X-ray luminosity ratio was a factor
  of $\sim12$ higher in the early Universe compared to today.  We
  explore possible explanations for this dramatic evolution and
  demonstrate how forthcoming {\it Herschel} observations will
  discriminate between these scenarios and, while doing so, identify
  potential Compton-thick AGNs.  Summary of \citet{Mullaney10}.
\end{abstract}

\maketitle

%%%%%%%%%%%%%%%%%%%%%%%%%%%%%%%%%%%%%%%%%%%%
%% MAINMATTER
%%%%%%%%%%%%%%%%%%%%%%%%%%%%%%%%%%%%%%%%%%%%

\section{Introduction}
With their ability to penetrate large columns of gas and dust,
observations at both X-ray and infrared wavelengths are complementary
in the study of active galactic nuclei (hereafter, AGNs).  While
X-rays provide a largely uncontaminated view of the primary emission
process, measures of infrared emission yield detailed information
about the levels of dust and on-going star-formation surrounding AGNs.
Consequently, by combining results from both X-ray and infrared
observations, we can address some of the key remaining questions
concerning AGNs and their to links galaxy evolution, such as (1) the
properties of the obscuring AGN dusty torus and how it has evolved with
time, (2) the relative ratio between galaxy and black hole growth (i.e.,
star-formation and AGN activity) over cosmic time and (3) the numbers
of X-ray weak, infrared bright Compton-thick AGNs in the Universe.

Here, we provide a summary of our recently published investigation
into the mid to far-infrared (MIR [5-30~${\rm \mu m}$] and FIR
[30-300~${\rm \mu m}$], respectively) properties of X-ray detected
AGNs out to $z\sim2$ (see \citet{Mullaney10}).  As part of this study,
we develop new MIR/FIR diagnostics to identify those systems dominated
by AGN activity and derive their total infrared output (across
8-1000~${\rm \mu m}$; hereafter, $L_{\rm IR}$) from their 24 ~${\rm
  \mu m}$ and 70 ~${\rm \mu m}$ flux densities (hereafter, $S_{24}$
and $S_{70}$, respectively).  We then use these diagnostics to explore
how the infrared properties of X-ray detected AGNs vary as a function
of redshift and intrinsic power (assumed to correlate with X-ray
luminosity, $L_{\rm X}$).  In doing so, we find that the average
infrared to X-ray luminosity ratio ($L_{\rm IR}/L_{\rm X}$) of
moderate luminosity AGNs has evolved strongly over cosmic time.
Finally, we explore how these diagnostics will be combined with the
results from planned {\it Herschel} surveys to further enhance our
understanding of AGNs and their role in galaxy evolution.

\section{MIR/FIR AGN Diagnostics}
\begin{figure}
  \includegraphics[height=6cm]{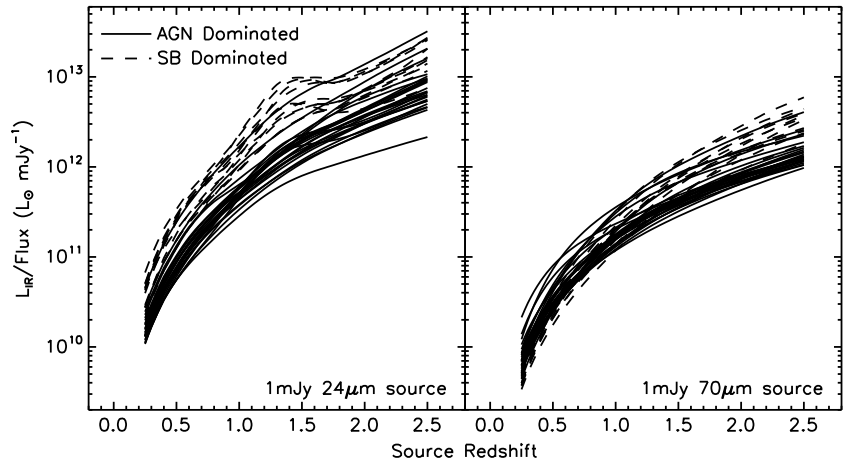}
  \caption{$L_{\rm IR}$/$S_{24}$ and $L_{\rm IR}$/$S_{70}$ ratios
    derived from our local sample of AGNs as a function of redshift,
    derived from our local sample of AGNs and assuming $S_{24}=1$~mJy
    (left) and $S_{70}=1$~mJy.  Each track corresponds to a different
    AGN from our local sample.  Dashed lines correspond to those AGNs
    whose infrared emission is dominated by star-formation activity,
    whereas solid lines represent those classed as having
    AGN-dominated infrared SEDs.  For all redshifts considered,
    70~${\rm \mu m}$ flux densities provide a significantly more
    precise measure of the total AGN infrared luminosity, displaying a
    smaller spread among all tracks.}

\end{figure}

We use a sample of 36 local (i.e., $z<0.1$) AGNs to (1) quantify the
$L_{\rm IR}$/$S_{24}$ and $L_{\rm IR}$/$S_{70}$ ratios as a function
of redshift and (2) explore how the $S_{70}$/$S_{24}$ flux ratio
correlates with the relative AGN contribution to the infrared output.
To construct this sample, we selected all AGNs from the {\it
  Swift}-BAT catalogue that have well sampled 5 - 100~${\rm \mu m}$
(observed) infrared SEDs, incorporating both archival {\it
  Spitzer}-IRS spectra and {\it IRAS} flux densities.  By selecting
from the {\it Swift}-BAT catalogue, we ensure that the X-ray
properties of this local sample are representative of our main, high
redshift sample of AGNs (i.e. $L_{\rm X} \sim 10^{41}-10^{45}~{\rm erg
  s^{-1}}$, $N_{\rm H} > 10^{20}~{\rm cm^{-2}}$
;\cite{Alexander03,Winter09}).  Using the 6.2 um PAH emission feature
as an indicator of star formation activity \cite{Genzel98}, we
identify those AGNs whose infrared output is dominated by
starbursts. The remainder are classified as AGN-dominated systems.

In Fig. 1 we present the $L_{\rm IR}$/$S_{24}$ and $L_{\rm
  IR}$/$S_{70}$ ratios derived from our sample of local, X-ray
detected AGNs.  Each track in these plots is derived from an
individual AGN's infrared SED, classified in terms of their dominant
source of infrared emission (i.e., AGN or starburst).  At all
redshifts considered, a given $S_{70}$ flux density (in this case,
1~mJy) corresponds to a significantly smaller range in $L_{\rm IR}$
than could be derived from $S_{24}$ alone.  For example, estimates of
$L_{\rm IR}$ for a $z=1.5$ source based solely on $S_{24}$ can span
over an order of magnitude (i.e., a factor of $\sim12$).  On the other
hand, an equivalent source measured solely at 70~${\rm \mu m}$ would
have its $L_{\rm IR}$ constrained to within a factor of $\sim4$.  We
note these quoted ranges of $L_{\rm IR}$ are derived from our full
sample of low redshift AGNs, irrespective of infrared classification
(i.e., AGN or starburst-dominated).

By classifying the local X-ray detected AGNs in terms of their
dominant source of infrared emission, we can investigate whether the
$S_{70}$/$S_{24}$ flux ratio can be used to identify AGN-dominated
systems.  This seems plausible based on evidence that, compared to
inactive galaxies, AGNs show an excess of MIR emission produced by
dust heated by the intense radiation field of the central engine
\cite{Grijp87}.  By exploiting the well constrained SEDs of our local
AGN sample, we find that this excess results in AGN-dominated systems
having lower $S_{70}$/$S_{24}$ ratios compared to their
starburst-dominated counterparts, with an AGN/starburst division
around $S_{70}$/$S_{24} \sim 5$.  This diagnostic remains valid for
sources at redshifts $<1.5$, beyond which longer wavelength data can
be used to discriminate between AGN and starburst-dominated systems
(see \S4.3 and Fig. 8 of \citealt{Mullaney10}).

\begin{figure}
  \includegraphics[height=7cm]{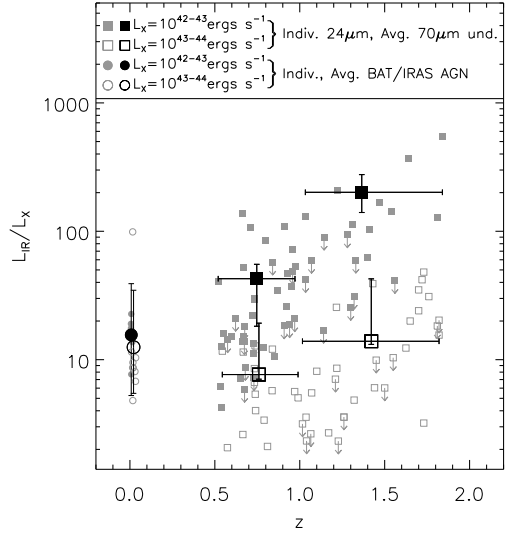}
  \caption{$L_{\rm IR}/L_{X}$ versus redshift for the X-ray detected
    AGNs in our high and low redshift samples (circles and squares,
    respectively).  Points are filled according to X-ray luminosity
    (see key).  Vertical error bars indicate the range of $L_{\rm
      IR}/L_{X}$ produced by assuming the various tracks in Fig. 1.
    Large, black symbols indicate the results derived from stacking
    analysis of the 70~${\rm \mu m}$ observations while small, grey
    squares refer to results derived from individual $24~{\rm \mu m}$
    flux densities (upper limits assume a $24~{\rm \mu m}$ limiting
    sensitivity of $50~\mu {\rm Jy}$; see \citet{Mullaney10} for
    details).  Both the stacking analysis and the individual $24~{\rm
      \mu m}$ detections clearly show a consistent increase in the
    $L_{\rm IR}/L_{X}$ ratio of $L_{\rm X}=10^{42-43}~{\rm erg
      s^{-1}}$ AGN from $z\approx0$ to $z=1-2$. From
    \citet{Mullaney10}.}
\end{figure}

\section{The MIR-FIR Properties of X-ray AGNs}

By applying the above MIR/FIR diagnostics to the high redshift, X-ray
detected AGNs in the 1~Ms Chandra Deep Field South (hereafter, CDF-S),
we explore how the infrared properties of X-ray detected AGNs vary as
a function of redshift and intrinsic power (i.e., $L_{\rm X}$).  The
intrinsic X-ray luminosities and absorbing column densities of the
CDF-S AGNs were taken from \citet{Tozzi06}.  The observed 24~${\rm \mu
  m}$ and 70~${\rm \mu m}$ flux densities of the CDF-S AGNs were
obtained from the FIDEL and GOODS {\it Spitzer} legacy surveys.
However, as only $\sim$10\% of the X-ray AGNs in this field are
formally detected at 70~${\rm \mu m}$, we relied on stacking analysis
to gain insight into their average infrared properties.

\subsection{The Strongly Evolving Infrared Luminosities of AGNs}

In Fig. 2 we plot the $L_{\rm IR}/L_{\rm X}$ luminosity ratios of the
X-ray AGNs as a function of redshift and separated in terms of $L_{\rm
  X}$.  We include in this plot results derived from both our local
(i.e., {\it Swift}-BAT) and high redshift (i.e., CDF-S) samples of
X-ray detected AGNs.  It is clear from this plot that high redshift,
moderate X-ray luminosity AGNs (i.e., defined here as $L_{\rm X} =
10^{42}-10^{43}~{\rm erg s^{-1}}$) are significantly more {\it
  infrared} luminous (per unit accretion power output) than their low
redshift counterparts.  Indeed, we note a factor of
$12.7^{+7.1}_{-2.6}$ increase in the $L_{\rm IR}/L_{\rm X}$ luminosity
ratios between moderate $L_{\rm X}$ AGNs at $z\sim0$ and $z=1-2$.
However, in the same redshift interval we measure no significant
change in the $L_{\rm IR}/L_{\rm X}$ luminosity ratios of the most
X-ray luminous AGNs in our sample (i.e., with $L_{\rm X} =
10^{43}-10^{44}~{\rm erg s^{-1}}$).

Clearly, results based solely on stacking analysis can be subject to
significant systematic uncertainties, not least from the possibility
that stacked fluxes can be dominated by a small number of bright
sources lying just below the detection threshold of the survey.  To
check whether such bright, yet undetected, sources are responsible for
the observed increase in the $L_{\rm IR}/L_{\rm X}$ luminosity ratios,
we also present in Fig. 2 points derived from the 24~${\rm \mu m}$
flux densities of the individual X-ray AGNs.  Although measures of
$L_{\rm IR}$ derived from $S_{24}$ alone are subject to significant
systematic uncertainties (as outlines above), as over 70\% of the
CDF-S AGNs are detected in at 24 um, such estimates provide an
alternative and independent approach to confirming trends derived from
the 70~${\rm \mu m}$ stacks.  We note that those moderate X-ray
luminosity AGNs individually detected at 24 um same follow the trend
of increasing $L_{\rm IR}/L_{\rm X}$ luminosity ratio with redshift
and are not dominated by small numbers of bright, individual sources.

In summary, both 24~${\rm \mu m}$ and 70~${\rm \mu m}$ data reveal a
strongly evolving $L_{\rm IR}/L_{\rm X}$ luminosity ratio among
moderate luminosity X-ray AGNs, but there is no evidence of any change among
the most luminous X-ray AGNs in our sample.

\subsection{What is Driving the Observed Increase in $L_{\rm IR}$?}
The evolution in $L_{\rm IR}/L_{\rm X}$ for the moderate luminosity
AGNs could be due to changes in the AGN dusty torus or increased
star-formation in the host galaxies.  However, as so few of the CDF-S
X-ray AGNs are detected at 70~${\rm \mu m}$, and the stacked
$S_{70}/S_{24}$ ratios lie around the division between AGN and
starburst-dominated systems (i.e., $S_{70}/S_{24}\sim5$), we cannot
currently use this diagnostics to distinguish between these scenarios.
Because all other major discriminators between AGN activity and star
formation (e.g., from {\it Spitzer}-IRAC photometry; \citealt{Stern05})
also provide inconclusive results for the dominant source of infrared
emission in these AGNs, we consider the impact that {\it each} of
these scenarios would have on our understanding of AGNs and the
influence they have on their host galaxies.

If attributed to star-formation alone, the observed increase in the
$L_{\rm IR}/L_{\rm X}$ luminosity ratio would imply a significantly
enhanced ratio of galaxy to black hole growth in the early Universe.
If all this star-formation were to take place within the bulge of the
host galaxy, then this would have a significant impact on the
development of black-hole/bulge relationships observed in the local
Universe.  Using star formation rates derived from
\citet{Kennicutt98}, we estimate that the ratio between star-formation
and mass accretion ($\dot{M}_{\rm SF}/\dot{M}_{\rm Acc}\sim 500$) is
comparable to today's BH-bulge mass ratios ($M_{\rm Bulge}/M_{\rm
  BH}\sim800$).  This could be interpreted as tentative evidence that
the links between SMBHs and their host bulges were initially forged at
these early times.  We note, however, that we have not factored in the
growth of inactive galaxies in this assessment.

\begin{figure}
  \includegraphics[height=7cm]{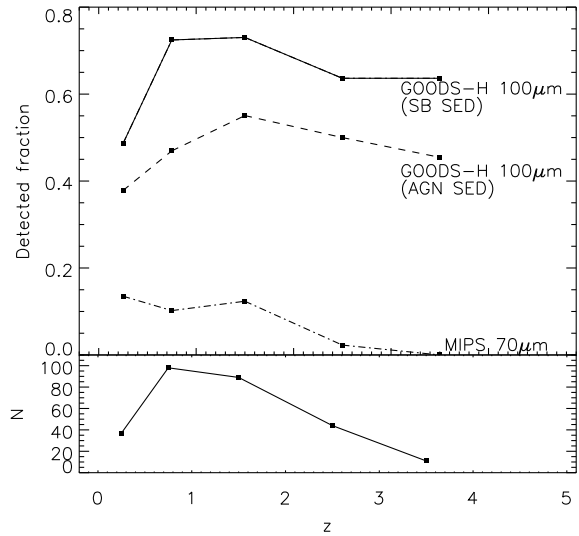}
  \caption{\textit{Top:} Predicted fractions of CDF-S X-ray AGNs
    detected in the upcoming GOODS-H ultra-deep infrared survey to be
    undertaken by the \textit{Herschel Telescope}.  The expected
    observed $100~{\rm \mu m}$ flux densities are based on $24~{\rm
      \mu m}$ flux densities of the CDF-S AGNs extrapolated along the
    average starburst-dominated and AGN-dominated SEDs, derived from
    our local sample of AGNs.  If all the X-ray AGNs were to have
    starburst-dominated IR SEDs, we would expect to detect {\it at
      least} as high a fraction at $100~{\rm \mu m}$ as we currently
    detect at $24~{\rm \mu m}$.  For comparison we include the
    fractions of AGNs currently detected at $70~{\rm \mu m}$.
    \textit{Bottom:} The total number of CDF-S X-ray AGNs in each
    bin. From \citet{Mullaney10}.}
\end{figure}

On the other hand, the increase in the $L_{\rm IR}/L_{\rm X}$
luminosity ratio could also be driven by changes associated with the
AGN itself.  As the increase is observed in an X-ray luminosity
matched sample of AGNs, this change is unlikely to be caused by
differences in the power output of the central engine.  On the other
hand, an increase in the dust covering factor at high redshifts would
manifest itself as an increase in the average infrared luminosity.
This interpretation agrees, in principle, with results derived from
deep X-ray observations that suggest that AGNs in the early Universe
were, on average, more heavily obscured than present- a result also
supported by evolutionary models of AGNs
(e.g. \citealt{Hasinger08,Hopkins06}).  However, the increase in dust
covering factor needed to account for the observed increase in $L_{\rm
  IR}$ is significantly higher than suggested by those studies.
Indeed, if one assumes a typical dust covering factor of $\sim40\%$ in
the local Universe, then a simplistic interpretation of our results
would indicate dust covering factors well in excess of 100\% in the
early Universe.  We note however, that this assumes a linear
relationship between $L_{\rm IR}$ and dust covering factor when it is
far from clear that this is indeed the case.

Finally, as the measured infrared luminosities of the high-z X-ray
AGNs are based solely on extrapolation using the SEDs of local AGNs,
it is plausible that the apparent increase in $L_{\rm IR}$ is a result
of systematic changes in the SEDs of AGNs across the redshifts
considered.  Indeed, it has been suggested that the SEDs of purely
star-forming galaxies have undergone some degree of change since
$z\sim2$ (e.g., Magnelli et al., {\it in prep.}).  Perhaps this could
be also true for AGNs? However, with so few high redshift X-ray AGNs
detected at far-infrared wavelengths, it is extremely difficult to
determine whether this is indeed the case.  Future deep surveys to be
undertaken by {\it Herschel} promise to remedy this situation and
distinguish between these different scenarios.

Irrespective of the cause of the increased infrared luminosity, a
thorough understanding of how $L_{\rm IR}/L_{\rm X}$ evolves will be
necessary in order to identify potential X-ray weak, infrared bright
AGNs at high redshifts and determine their intrinsic luminosities.

\section{Predictions for Deep Herschel Observations}
With its vastly superior sensitivity at far-infrared wavelengths, the
recently launched {\it Herschel Observatory} will probe a whole new
region of the $L_{\rm IR}-z$ parameter space.  Our estimates (based on
extrapolation of the 24~${\rm \mu m}$ flux densities) indicate that as
many as 70\% of the X-ray detected AGNs in the CDF-S will be detected
at 100~${\rm \mu m}$ in the planned ultra-deep {\it Herschel}
observations of this field (i.e., GOODS-H, P.I.: D. Elbaz; see
Fig. 3), compared to the $\sim$10\% detected in the deepest {\it
  Spitzer} FIR surveys.  With such large numbers of detections, we
will be able to constrain the far-infrared SEDs of the majority of the
X-ray AGNs in the deep fields, providing clear insights into the
dominant source of infrared emission in these objects (i.e., AGN or
star-formation activity).  This, in turn, will enable us to establish
the full extent and cause of the observed increase in the $L_{\rm
  IR}/L_{\rm X}$ luminosity ratio.  Furthermore, the diagnostics
described here and in \citet{Mullaney10} provide the means to use {\it
  Herschel} photometry to identify significant numbers of
AGN-dominated systems, including those X-ray weak, Compton-thick AGNs
currently missed at all other wavelengths.

%%%%%%%%%%%%%%%%%%%%%%%%%%%%%%%%%%%%%%%%%%%%%%%%
%% BACKMATTER
%%%%%%%%%%%%%%%%%%%%%%%%%%%%%%%%%%%%%%%%%%%%%%%%

% \begin{theacknowledgments}
%   JRM and DMA are funded by The Leverhulme Trust.  DMA is also funded
%   by The Royal Society.
% \end{theacknowledgments}



% \bibliography{List}

\begin{thebibliography}{10}
\expandafter\ifx\csname natexlab\endcsname\relax\def\natexlab#1{#1}\fi
\providecommand{\enquote}[1]{``#1''}
\expandafter\ifx\csname url\endcsname\relax
  \def\url#1{\texttt{#1}}\fi
\expandafter\ifx\csname urlprefix\endcsname\relax\def\urlprefix{URL }\fi
\providecommand{\eprint}[2][]{\url{#2}}

\bibitem[{Mullaney} et~al.(2010)]{Mullaney10}
J.~R. {Mullaney} et al., \emph{MNRAS} \textbf{401}, 995--1012 (2010).

\bibitem[{Alexander} et~al.(2003)]{Alexander03}
D.~M. {Alexander} et al., \emph{AJ} \textbf{126}, 539--574 (2003).

\bibitem[{Winter} et~al.(2009)]{Winter09}
L.~M. {Winter} , R.~F. {Mushotzky}, C.~S. {Reynolds}, and J.~{Tueller},
  \emph{ApJ} \textbf{690}, 1322--1349 (2009).

\bibitem[{Genzel} et~al.(1998)]{Genzel98}
R.~{Genzel} et al., \emph{ApJ} \textbf{498},
  579--+ (1998).

\bibitem[{de Grijp} et~al.(1987)]{Grijp87}
M.~H.~K. {de Grijp}, J.~{Lub}, and G.~K. {Miley}, \emph{A\&ASS} \textbf{70},
  95--114 (1987).

\bibitem[{Tozzi} et~al.(2006)]{Tozzi06}
P.~{Tozzi} et al., \emph{A\&A}
  \textbf{451}, 457--474 (2006).

\bibitem[{Stern} et~al.(2005)]{Stern05}
D.~{Stern} et al., \emph{ApJ} \textbf{631}, 163--168 (2005).

\bibitem[{Kennicutt}(1998)]{Kennicutt98}
R.~C. {Kennicutt}, Jr., \emph{ARA\&A} \textbf{36}, 189--232 (1998).

\bibitem[{Hasinger}(2008)]{Hasinger08}
G.~{Hasinger}, \emph{A\&A} \textbf{490}, 905--922 (2008).

\bibitem[{Hopkins} et~al.(2006)]{Hopkins06}
P.~F. {Hopkins} et al., \emph{ApJS} \textbf{163}, 1--49 (2006).

\end{thebibliography}
\bibliographystyle{aipproc}   % if natbib is available

\end{document}